# Knightian Robustness from Regret Minimization

Alessandro Chiesa Silvio Micali Zeyuan Allen Zhu

MIT

April 1, 2014

**Abstract**

We consider auctions in which the players have very limited knowledge about their own valuations. Specifically, the only information that a *Knightian player* $i$ has about the profile of true valuations, $\theta^*$, consists of a set of distributions, from one of which $\theta_i^*$ has been drawn.

We analyze the social-welfare performance of the VCG mechanism, for unrestricted combinatorial auctions, when Knightian players that either

(a) choose a regret-minimizing strategy, or

(b) resort to regret minimization only to refine further their own sets of undominated strategies, if needed.

We prove that this performance is very good.

# 1 Introduction

In [CMZ14b] we motivate the problem of mechanism design for Knightian players, and prove that (1) dominant-strategy mechanisms for single-good and multi-unit auctions cannot provide good social-welfare efficiency, but (2) the second-price and Vickrey mechanisms deliver good social-welfare performance, for these two settings, in undominated strategies.

In this report, we prove that the VCG mechanism guarantees good social welfare in the presence of Knightian players who either (a) choose a regret-minimizing strategy, or (b) resort to regret minimization only to refine further their own sets of undominated strategies, if needed.

# 2 Model

We study *unrestricted combinatorial auctions*, where there are $n$ players and $m$ distinct goods. The set of possible allocations $\mathcal{A}$ consists of all possible partitions $A$ of $[m]$ into $1+n$ subsets, $A=(A_0,A_1,\ldots,A_n)$, where $A_0$ is the (possibly empty) set of unassigned goods and $A_i$ is the (possibly empty) set of goods assigned to player $i$.

For each player $i$, a valuation is a function mapping each possible subset of the goods to a non-negative real, and the set of all possible valuations is $\Theta_i=\{\theta_i : 2^{[m]} \to \mathbb{R}_{\geq 0} \mid \theta_i(\varnothing)=0\}$. The profile of the players' true valuations is $\theta^*=(\theta_1^*,\ldots,\theta_n^*)\in\Theta$.

The set of possible outcomes is $\Omega \stackrel{\text{def}}{=} \mathcal{A}\times\mathbb{R}_{\geq 0}^n$. If $(A,P)\in\Omega$, we refer $P_i$ as the price charged to player $i$. We assume quasi-linear utilities. That is, the utility function $U_i$ of a player $i$ maps a valuation $\theta_i$ and an outcome $\omega=(A,P)$ to $U_i(\theta_i,\omega) \stackrel{\text{def}}{=} \theta_i(A_i)-P_i$.

If $\omega$ is a distribution over outcomes, we also denote by $U_i(\theta_i,\omega)$ the expected utility of player $i$.

## 2.1 Knightian Valuation Uncertainty

In our model, a player $i$'s sole information about $\theta^*$ consists of $\mathcal{K}_i$, a set of distributions over $\Theta_i$, from one of which $\theta_i^*$ has been drawn. (The true valuations are uncorrelated.) That is, $\mathcal{K}_i$ is $i$'s sole (and private) information about his own true valuation $\theta_i^*$.

Furthermore, for every opponent $j$, $i$ has no information (or beliefs) about $\theta_j^*$ or $\mathcal{K}_j$.

Given that all he cares about is his expected (quasi-linear) utility, a player $i$ may 'collapse' each distribution $D_i \in \mathcal{K}_i$ to its expectation $\mathbb{E}_{\theta_i \sim D_i}[\theta_i]$.[1] Therefore, for unrestricted combinatorial auctions, a *mathematically equivalent* formulation of the Knightian valuation model is the following:

**Definition 2.1** (Knightian valuation model)**.** *For each player $i$, $i$'s sole information about $\theta^*$ is a set $K_i$, the* candidate (valuation) set of $i$*, such that $\theta_i^* \in K_i \subset \Theta_i$.*

*We refer to an element of $K_i$ as a* candidate valuation*.*

In Knightian valuation model, a mechanism's performance will of course depend on the inaccuracy of the players' candidate sets, which we measure as follows.

**Definition 2.2.** *The candidate set $K_i$ of a player $i$ is* (at most) $\delta$-approximate *if, for each subset $S \subseteq [m]$, letting $K_i(S) \stackrel{\text{def}}{=} \{\theta_i(S) \mid \theta_i \in K_i\}$, $\sup K_i(S) - \inf K_i(S) \leq \delta$.*

*An auction is* (at most) $\delta$-approximate *if each $K_i$ is $\delta$-approximate.*

## 2.2 Social Welfare, Mechanisms, and Knightian Dominance

**Social welfare.** The social welfare of an allocation $A = (A_0, A_1, \ldots, A_n)$, $\mathrm{SW}(A)$, is defined to be $\sum_i \theta_i^*(A_i)$; and the maximum social welfare, MSW, is defined to be $\max_{A \in \mathcal{A}} \mathrm{SW}(A)$. (That is, SW and MSW continue to be defined relative to the players' true valuations $\theta_i^*$, whether or not the players know them exactly.)

More generally, the social welfare of an allocation $A$ relative to a valuation profile $\theta$, $\mathrm{SW}(\theta, A)$, is $\sum_i \theta_i(A_i)$; and the maximum social welfare relative to $\theta$, $\mathrm{MSW}(\theta)$, is $\max_{A \in \mathcal{A}} \mathrm{SW}(\theta, A)$. Thus, $\mathrm{SW}(A) = \mathrm{SW}(\theta^*, A)$ and $\mathrm{MSW} = \mathrm{MSW}(\theta^*)$.

**Mechanisms and strategies.** A mechanism $M$ specifies, for each player $i$, a set $S_i$. We interchangeably refer to each member of $S_i$ as a pure *strategy/action/report* of $i$, and similarly, a member of $\Delta(S_i)$ a mixed strategy/action/report of $i$.

After each player $i$, simultaneously with his opponents, reports a strategy $s_i$ in $S_i$, $M$ maps the reported strategy profile $s$ to an outcome $M(s) \in \Omega$.

[1] Whatever the auction mechanism used, this equivalence holds for any auction where each $\Theta_i$ is a *convex* set. In particular, this includes unrestricted combinatorial auctions of $m$ distinct goods.

If $M$ is probabilistic, then $M(s) \in \Delta(\Omega)$. Thus, as per our notation, $U_i(\theta_i, M(s)) \stackrel{\text{def}}{=} \mathbb{E}_{\omega \sim M(s)}[U_i(\theta_i, \omega)]$ for each player $i$.

Note that $S_i = \Theta_i$ for the direct mechanisms in the classical setting.

**The VCG mechanism.** In our auctions, the VCG mechanism, denoted $\mathsf{VCG}$, maps a profile of valuations $\theta \in \Theta_1 \times \cdots \times \Theta_n$, to an outcome $(A, P)$, where

$A \in \arg\max_{A \in \mathcal{A}} \mathrm{SW}(\theta, A)$ and, for each player $i$, $P_i = \mathrm{MSW}(\theta_{-i}) - \sum_{j \neq i} \theta(A_i)$. Ties are broken by preferring subsets with smaller cardinalities.[2]

**Knightian regret-minimizing strategies.** Given a mechanism $M$, the (maximum) regret of a pure strategy $s_i$ of a player $i$ with candidate set $K_i$ is

$$R_i(K_i, s_i) \stackrel{\text{def}}{=} \max_{\theta_i \in K_i} \max_{s_{-i}} \left( \max_{s'_i} U_i\big(\theta_i, M(s'_i, s_{-i})\big) - U_i\big(\theta_i, M(s_i, s_{-i})\big) \right) .$$

A pure strategy $s_i$ is *regret-minimizing* among all pure strategies of a player $i$ with a candidate set $K_i$, in symbols $s_i \in \mathsf{RM}_i^{\mathsf{pure}}(K_i)$, if $R_i(K_i, s_i) \geq R_i(K_i, s'_i)$ for all other pure strategies $s'_i$ of $i$. We let $\mathsf{RM}^{\mathsf{pure}}(K) \stackrel{\text{def}}{=} \mathsf{RM}_1^{\mathsf{pure}}(K_1) \times \cdots \times \mathsf{RM}_n^{\mathsf{pure}}(K_n)$.

When allowing mixed strategies, the (expected) regret of a (possibly mixed) strategy $\sigma_i$ of a player $i$ with candidate set $K_i$ is

$$R_i(K_i, \sigma_i) \stackrel{\text{def}}{=} \max_{\theta_i \in K_i} \max_{s_{-i}} \left( \max_{s'_i} U_i\big(\theta_i, M(s'_i, s_{-i})\big) - \mathbb{E}_{s_i \sim \sigma_i} U_i\big(\theta_i, M(s_i, s_{-i})\big) \right) .$$

We similarly define $\mathsf{RM}_i^{\mathsf{mix}}(K_i)$ as the set of strategies of a player $i$ that minimize regret among all mixed strategies, and let $\mathsf{RM}^{\mathsf{mix}}(K) \stackrel{\text{def}}{=} \mathsf{RM}_1^{\mathsf{mix}}(K_1) \times \cdots \times \mathsf{RM}_n^{\mathsf{mix}}(K_n)$.

# 3 Result

In $\delta$-approximate combinatorial auctions with $n$ players and $m$ goods, the VCG guarantees social welfare $\geq \mathrm{MSW} - 2\min\{n, m\}\delta$ in *pure* regret-minimizing strategies:

[2] If giving subsets $A$ or $B \subsetneq A$ to player $i$ provides the same social welfare, then the VCG will give $B$ to player $i$.

**Theorem 1.** *In a combinatorial Knightian auction with $n$ players and $m$ goods, for all $\delta$, all products $K$ of $\delta$-approximate candidate sets, all profiles $\theta \in K$, and all profiles of strategies $v \in \mathsf{RM}^{\mathsf{pure}}(K)$, it holds that*

$$\mathrm{SW}(\theta, \mathsf{VCG}(v)) \geq \mathrm{MSW}(\theta) - 2\min\{m, n\}\delta \ .$$

**Discussion.** Theorem 1 says that, in combinatorial Knightian auctions, the performance of the VCG in (pure) regret minimizing strategies is very good. Moreover, because of the result proved in [CMZ14a], the same holds for when a player resorts to regret minimization only to refine further his own sets of undominated strategies.[3]

Theorem 1 is less intuitive than it seems, because in a combinatorial, Knightian, VCG auction it is not obvious which strategies are regret-minimizing. Consider a player $i$ who (1) happens to know that his true valuation for some subset of the good $S$ lies in some interval $[x_S, x_S + \delta]$, and (2) chooses to play a pure, regret-minimizing strategy $v_i$. At first glance, it would appear that $v_i(S)$ should coincide with the center of the interval, that is, $v_i(S) = x_S + \delta/2$. In reality, however, $v_i(S)$ need not even belong to the interval $[x_S, x_S + \delta]$. Nevertheless, we prove that it cannot lie too far from the interval.

We would like to mention that Theorem 1 continues to hold when *mixed* regret-minimizing strategies are allowed, but with a worse bound. Roughly, $\min\{n, m\}$ is replaced by $n^2$ (or even $n \log n$ if the valuations are set-monotone).[4]

**Proof.** We begin by noting that, because the VCG is dominant-strategy-truthful in the exact-valuation model, the (maximum) regret of a pure strategy $v_i$ of a player $i$ with candidate set $K_i$ in the VCG mechanism becomes

$$\begin{aligned} R_i(K_i, v_i) &\stackrel{\text{def}}{=} \max_{\theta_i \in K_i} \max_{v_{-i}} \Big( \max_{v'_i} U_i\big(\theta_i, \mathsf{VCG}(v'_i, v_{-i})\big) - U_i\big(\theta_i, \mathsf{VCG}(v_i, v_{-i})\big) \Big) \\ &= \max_{\theta_i \in K_i} \max_{v_{-i}} \Big( U_i\big(\theta_i, \mathsf{VCG}(\theta_i, v_{-i})\big) - U_i\big(\theta_i, \mathsf{VCG}(v_i, v_{-i})\big) \Big) \ , \end{aligned}$$

---

[3]A pure strategy $s_i$ of a player $i$ with a candidate set $K_i$ is *(weakly) undominated*, in symbols $s_i \in \mathsf{UD}_i(K_i)$, if $i$ does not have another (possibly mixed) strategy $\sigma_i$ such that
(1) $\forall \theta_i \in K_i \ \forall s_{-i} \in S_{-i} \quad \mathbb{E} U_i\big(\theta_i, M(\sigma_i, s_{-i})\big) \geq U_i\big(\theta_i, M(s_i, s_{-i})\big)$, and
(2) $\exists \theta_i \in K_i \ \exists s_{-i} \in S_{-i} \quad \mathbb{E} U_i\big(\theta_i, M(\sigma_i, s_{-i})\big) > U_i\big(\theta_i, M(s_i, s_{-i})\big)$.

[4]That is, $v_i(S) \leq v_i(T)$ for all $S \subseteq T \subseteq [m]$, all $i$, and all $v_i \in \Theta_i$.

Moreover, by the very definition of the VCG, we have

$$U_i\big(\theta_i, \mathsf{VCG}(v_i, v_{-i})\big) = \mathrm{SW}\big((\theta_i, v_{-i}), \mathsf{VCG}(v_i, v_{-i})\big) - \mathrm{MSW}(v_{-i}) \ .[5]$$

Therefore in the VCG case, we can further simplify the definition of regret as follows:

$$\begin{aligned} R_i(K_i, v_i) &= \max_{\theta_i \in K_i} \max_{v_{-i}} \Big( \mathrm{SW}\big((\theta_i, v_{-i}), \mathsf{VCG}(\theta_i, v_{-i})\big) - \mathrm{SW}\big((\theta_i, v_{-i}), \mathsf{VCG}(v_i, v_{-i})\big)\Big) \\ &= \max_{\theta_i \in K_i} \max_{v_{-i}} \Big( \mathrm{MSW}\big(\theta_i, v_{-i}\big) - \mathrm{SW}\big((\theta_i, v_{-i}), \mathsf{VCG}(v_i, v_{-i})\big)\Big) \ . \qquad (3.1) \end{aligned}$$

For each player $i$, each candidate set $K_i \subset \Theta_i$, and each subset $T \subseteq [m]$, we let

$$K_i(T) \stackrel{\text{def}}{=} \{\theta_i(T)\}_{\theta_i \in K_i}, \quad K_i^{\perp}(T) \stackrel{\text{def}}{=} \inf K_i(T),$$
$$K_i^{\top}(T) \stackrel{\text{def}}{=} \sup K_i(T), \quad K_i^{\text{mid}}(T) \stackrel{\text{def}}{=} (K_i^{\perp}(T) + K_i^{\top}(T))/2 \ .$$

To prove Theorem 1, we rely on two intermediate claims. The first one identifies, for every player $i$, a strategy $v_i$ with regret no larger than $\delta$.

**Claim 3.1.** *For every player $i$, let $v_i^*(T) \stackrel{\text{def}}{=} K_i^{\text{mid}}(T)$ for each $T \subseteq [m]$. Then $R_i(K_i, v_i^*) \leq \delta$.*

*Proof of Claim 3.1.* According to the first equality of (3.1), it suffices to show that

$$\forall \theta_i \in K_i \ \forall v_{-i}, \quad \mathrm{SW}\big((\theta_i, v_{-i}), \mathsf{VCG}(\theta_i, v_{-i})\big) - \mathrm{SW}\big((\theta_i, v_{-i}), \mathsf{VCG}(v_i^*, v_{-i})\big) \leq \delta \ .$$

Let $\omega_1 = \mathsf{VCG}(\theta_i, v_{-i})$ and $\omega_2 = \mathsf{VCG}(v_i^*, v_{-i})$.

Recall that, in a combinatorial auction, a valuation $\theta_i \in \Theta_i$ of player $i$ maps subsets of $[m]$ to $\mathbb{R}_{\geq 0}$. For convenience, we extend $\theta_i$ to map an outcome $\omega = (A, P)$ to $\mathbb{R}_{\geq 0}$ as follows: $\theta_i(\omega) \stackrel{\text{def}}{=} \theta_i(A_i)$.

Under this notation, we have $v_i^*(\omega_2) + v_{-i}(\omega_2) \geq v_i^*(\omega_1) + v_{-i}(\omega_1)$, because the VCG maximizes social welfare relative to the strategy profile $(v_i^*, v_{-i})$. Using this inequality, we deduce that

$$\begin{aligned} &\mathrm{SW}\big((\theta_i, v_{-i}), \mathsf{VCG}(\theta_i, v_{-i})\big) - \mathrm{SW}\big((\theta_i, v_{-i}), \mathsf{VCG}(v_i^*, v_{-i})\big) \\ = \ &\big(\theta_i(\omega_1) + v_{-i}(\omega_1)\big) - \big(\theta_i(\omega_2) + v_{-i}(\omega_2)\big) \\ = \ &\big(\theta_i(\omega_1) - \theta_i(\omega_2)\big) + \big(v_{-i}(\omega_1) - v_{-i}(\omega_2)\big) \end{aligned}$$

[5] This is because, suppose that the VCG mechanism picks an outcome $\omega = \mathsf{VCG}(v_i, v_{-i})$, allocating player $i$ subset $A_i$ and others $A_{-i}$. Then, $i$'s price is $\mathrm{MSW}(v_{-i}) - v_{-i}(A_{-i})$ in $\omega$. This induces a total utility of $\theta_i(A_i) + v_{-i}(A_{-i}) - \mathrm{MSW}(v_{-i}) = \mathrm{SW}((\theta_i, v_{-i}), \omega) - \mathrm{MSW}(v_{-i})$.

$$\leq \big(\theta_i(\omega_1) - \theta_i(\omega_2)\big) + \big(v_i^*(\omega_2) - v_i^*(\omega_1)\big) \ .$$

Suppose player $i$ gets subset $T_1 \subseteq [m]$ in outcome $\omega_1$, and subset $T_2 \subseteq [m]$ in outcome $\omega_2$. Then

$$\begin{aligned}
&\big(\theta_i(\omega_1) - \theta_i(\omega_2)\big) + \big(v_i^*(\omega_2) - v_i^*(\omega_1)\big) \\
=\ & \big(\theta_i(T_1) - v_i^*(T_1)\big) + \big(v_i^*(T_2) - \theta_i(T_2)\big) \\
\leq\ & K_i^{\top}(T_1) - K_i^{\text{mid}}(T_1) + K_i^{\text{mid}}(T_2) - K_i^{\perp}(T_2) \\
\leq\ & \frac{\delta}{2} + \frac{\delta}{2} = \delta \ . \qquad \square
\end{aligned}$$

Let us now prove another claim.

**Claim 3.2.** *Let $v_i$ be any strategy of player $i$ such that $R_i(K_i, v_i) \leq \delta$. Then:*

*(a) for every $T \subseteq [m]$:*

$$K_i^{\text{mid}}(T) - \max_{T' \subseteq T} v_i(T') \leq \delta - \frac{K_i^{\top}(T) - K_i^{\perp}(T)}{2} \ , \ \textit{and}$$

*(b) for every $T \subseteq [m]$ such that $v_i(T) > v_i(T')$ for all $T' \subsetneq T$:*

$$|v_i(T) - K_i^{\text{mid}}(T)| \leq \delta - \frac{K_i^{\top}(T) - K_i^{\perp}(T)}{2} \ .$$

*Proof.* Since the case of $T = \varnothing$ is trivial, we assume below that $T \neq \varnothing$. We first prove part (a).

Suppose that (a) is not true. Then, there exists $T$ such that

$$K_i^{\text{mid}}(T) - \max_{T' \subseteq T} v_i(T') > \delta - \frac{K_i^{\top}(T) - K_i^{\perp}(T)}{2} \ . \tag{3.2}$$

We contradict our assumption on $v_i$ by showing that $R_i(K_i, v_i) > \delta$.

To show $R_i(K_i, v_i) > \delta$, as per (3.1), we must find some $v_{-i}$ and some $\theta_i$ so that

$$\text{MSW}\big(\theta_i, v_{-i}\big) - \text{SW}\big((\theta_i, v_{-i}), \mathsf{VCG}(v_i, v_{-i})\big) > \delta \ . \tag{3.3}$$

Let $j$ be an arbitrary player other than $i$. We choose $\theta_i \in K_i$ such that $\theta_i(T) =$

$K_i^\top(T)$,[6] and $v_{-i}$ as follows: for every $S \subseteq [m]$

$$v_j(S) \stackrel{\text{def}}{=} \begin{cases} H & \text{if } S = \overline{T} \\ H + \varepsilon + \max_{T' \subseteq T} v_i(T') & \text{if } S = [m] \\ 0 & \text{otherwise} \end{cases} \quad \text{and} \quad v_k(S) \stackrel{\text{def}}{=} 0 \text{ for every } k \notin \{i, j\}.$$

Above, $\varepsilon > 0$ is some sufficiently small real number, and $H$ is some huge real number (that is, $H$ is much bigger than $v_i(S)$ for any subset $S$).[7] It then is easy to verify that the outcome $\mathsf{VCG}(v_i, v_{-i})$ allocates $\varnothing$ to player $i$, and $[m]$ to player $j$. Therefore,

$$\mathrm{SW}\big((\theta_i, v_{-i}), \mathsf{VCG}(v_i, v_{-i})\big) = \theta_i(\varnothing) + v_j([m]) = H + \varepsilon + \max_{T' \subseteq T} v_i(T') \ .$$

On the other hand, $\mathrm{MSW}(\theta_i, v_{-i}) \geq \theta_i(T) + v_j(\overline{T}) = K_i^\top(T) + H$, and therefore

$$\mathrm{MSW}\big(\theta_i, v_{-i}\big) - \mathrm{SW}\big((\theta_i, v_{-i}), \mathsf{VCG}(v_i, v_{-i})\big) \geq \Big(K_i^\top(T) + H\Big) - \Big(H + \varepsilon + \max_{T' \subseteq T} v_i(T')\Big)$$

$$= K_i^\top(T) - \varepsilon - \max_{T' \subseteq T} v_i(T') = \frac{K_i^\top(T) - K_i^\perp(T)}{2} + K_i^{\mathrm{mid}}(T) - \varepsilon - \max_{T' \subseteq T} v_i(T') \ .$$

Finally, since $K_i^{\mathrm{mid}}(T) - \max_{T' \subseteq T} v_i(T')$ is strictly greater than $\delta - \frac{K_i^\top(T) - K_i^\perp(T)}{2}$, according to (3.2), there exists some sufficiently small $\varepsilon > 0$ to make $\frac{K_i^\top(T) - K_i^\perp(T)}{2} + K_i^{\mathrm{mid}}(T) - \varepsilon - \max_{T' \subseteq T} v_i(T') > \delta$. This proves (3.3) and concludes the proof of Claim 3.2a.

We now prove part Claim 3.2b.

One side of Claim 3.2b is easy: that is, $v_i(T) - K_i^{\mathrm{mid}}(T) \geq -(\delta - \frac{K_i^\top(T) - K_i^\perp(T)}{2})$. Indeed, this inequality follows from $\max_{T' \subseteq T} v_i(T') = v_i(T)$ and Claim 3.2a.

To show the other side, that is, $v_i(T) - K_i^{\mathrm{mid}}(T) \leq \delta - \frac{K_i^\top(T) - K_i^\perp(T)}{2}$, we again proceed by contradiction. Suppose there is some $T$ such that

$$v_i(T) - K_i^{\mathrm{mid}}(T) > \delta - \frac{K_i^\top(T) - K_i^\perp(T)}{2} \ . \tag{3.4}$$

We contradict our assumption on $v_i$ by showing that $R_i(K_i, v_i) > \delta$. Similarly to case (a), we need to find some $v_{-i}$ and some $\theta_i$ so that inequality (3.3) holds.

Let $j$ be an arbitrary player other than $i$. This time, we choose $\theta_i \in K_i$ such that

[6] Here we have implicitly assumed that $K_i^\top(T) = \sup K_i(T) = \max K_i(T)$, and thus we can pick $\theta_i \in K_i$ so that $\theta_i(T) = K_i^\top(T)$. If this is not the case, one can construct an infinite sequence $\theta_i^{(1)}, \theta_i^{(2)}, \cdots$ so that $\theta_i(T)$ approaches to $K_i^\top(T)$, and the rest of the proof remains unchanged.

[7] Notice that when $T = [m]$ we have $\overline{T} = \varnothing$ and one cannot assign $v_j(\varnothing)$ to be a nonzero number. In that case we can choose $H = 0$, and the rest of the proof still goes through.

$\theta_i(T) = K_i^{\perp}(T)$,[6] and choose $v_{-i}$ as follows: for every $S \subseteq [m]$

$$v_j(S) = \begin{cases} H & \text{if } S = \overline{T} \\ H - \varepsilon + v_i(T) & \text{if } S = [m] \\ 0 & \text{otherwise} \end{cases} \qquad \text{and} \qquad v_k(S) \stackrel{\text{def}}{=} 0 \text{ for every } k \notin \{i, j\}.$$

Again, $\varepsilon > 0$ is sufficiently small, and $H$ is huge.[7] It then is easy to verify that the outcome $\mathsf{VCG}(v_i, v_{-i})$ allocates $T$ to player $i$ and $\overline{T}$ to player $j$. Therefore,

$$\mathrm{SW}\big((\theta_i, v_{-i}), \mathsf{VCG}(v_i, v_{-i})\big) = \theta_i(T) + v_j(\overline{T}) = K_i^{\perp}(T) + H \ .$$

On the other hand, $\mathrm{MSW}(\theta_i, v_{-i}) \geq \theta_i(\varnothing) + v_j([m]) = H - \varepsilon + v_i(T)$. Therefore,

$$\begin{aligned} \mathrm{MSW}\big(\theta_i, v_{-i}\big) - \mathrm{SW}\big((\theta_i, v_{-i}), \mathsf{VCG}(v_i, v_{-i})\big) &\geq (H - \varepsilon + v_i(T)) - (K_i^{\perp}(T) + H) \\ &= v_i(T) - K_i^{\text{mid}}(T) + \frac{K_i^{\top}(T) - K_i^{\perp}(T)}{2} - \varepsilon \ . \end{aligned}$$

Finally, since $v_i(T) - K_i^{\text{mid}}(T)$ is strictly greater than $\delta - \frac{K_i^{\top}(T) - K_i^{\perp}(T)}{2}$ according to (3.4), there exists some sufficiently small $\varepsilon > 0$ to make $v_i(T) - K_i^{\text{mid}}(T) + \frac{K_i^{\top}(T) - K_i^{\perp}(T)}{2} - \varepsilon > \delta$. This proves (3.3) and concludes the proof of Claim 3.2b.

In sum, Claim 3.2 holds. □

Now we return to the proof of Theorem 1. Let $v = (v_1, \ldots, v_n) \in \mathsf{RM}^{\mathsf{pure}}(K)$ be a regret-minimizing pure strategy profile, and let $\theta \in K$ be a valuation profile.

For every player $i$, the strategy $v_i^*$ (i.e., the one reporting the 'middle points') has a regret at most $\delta$, owing to Claim 3.1. Since $v_i$ minimizes regret among all his strategies, we immediately have $R_i(K_i, v_i) \leq R_i(v_i^*, K_i) \leq \delta$. This shows that $v_i$ satisfies the initial hypothesis of Claim 3.2.

Now, letting $(A_0, A_1, \ldots, A_n)$ be the allocation in the outcome $\mathsf{VCG}(v_1, \ldots, v_n)$, we immediately have $v_i(A_i) \geq v_i(T')$ for any $T' \subsetneq A_i$ by the definition of the VCG. Furthermore, by our choice of the tie-breaking rule, this inequality must be strict: that is, $v_i(A_i) > v_i(T')$ for any $T' \subsetneq A_i$. Therefore, letting $T = A_i$, $T$ satisfies the hypothesis in Claim 3.2b. Thus, we conclude that

$$\begin{aligned} \forall i \in [n], \quad |v_i(A_i) - K_i^{\text{mid}}(A_i)| \leq \delta - \frac{K_i^{\top}(A_i) - K_i^{\perp}(A_i)}{2} &\leq \delta - |\theta_i(A_i) - K_i^{\text{mid}}(A_i)| \\ \implies |v_i(A_i) - \theta_i(A_i)| \leq \delta \ . \quad (3.5) \end{aligned}$$

Notice that, if $A_i = \varnothing$, then $v_i(\varnothing) = \theta_i(\varnothing) = 0$.

Next, letting $(B_0, B_1, \dots, B_n)$ be the allocation that maximizes the social welfare under $\theta$, we have

$$\sum_{i=1}^{n} v_i(A_i) \geq \sum_{i=1}^{n} \max_{T' \subseteq B_i} v_i(T') \tag{3.6}$$

because the VCG maximizes social welfare relative to $v = (v_1, \dots, v_n)$. Moreover, according to Claim 3.2a we have

$$\forall i \in [n], \quad K_i^{\text{mid}}(B_i) - \max_{T' \subseteq B_i} v_i(T') \leq \delta - \frac{K_i^{\top}(B_i) - K_i^{\perp}(B_i)}{2} \leq \delta - |\theta_i(B_i) - K_i^{\text{mid}}(B_i)|$$

$$\Longrightarrow \theta_i(B_i) - \max_{T' \subseteq B_i} v_i(T') \leq \delta \ . \tag{3.7}$$

Also notice that, if $B_i = \varnothing$, then $\theta_i(B_i) = \max_{T' \subseteq B_i} v_i(T') = 0$.

We are now ready to compute the social welfare guarantee.

$$\begin{aligned}
\mathrm{SW}(\theta, \mathsf{VCG}(v)) = \textstyle\sum_{i=1}^{n} \theta_i(A_i) \geq \ & \textstyle\sum_{i=1}^{n} v_i(A_i) - \sum_{i \in [n], A_i \neq \varnothing} \delta && \text{(using (3.5))} \\
\geq \ & \sum_{i=1}^{n} \max_{T' \subseteq B_i} v_i(T') - \sum_{i \in [n], A_i \neq \varnothing} \delta && \text{(using (3.6))} \\
\geq \ & \textstyle\sum_{i=1}^{n} \theta_i(B_i) - \sum_{i \in [n], A_i \neq \varnothing} \delta - \sum_{i \in [n], B_i \neq \varnothing} \delta && \text{(using (3.7))} \\
\geq \ & \mathrm{MSW}(\theta) - 2 \min\{n, m\} \delta \ .
\end{aligned}$$

This concludes the proof of Theorem 1. ■